\documentclass{elsart}

\usepackage{amssymb}
\usepackage{graphicx}% Include figure files

\begin{document}
\begin{frontmatter}

\title{The reflection of very cold neutrons from diamond powder nanoparticles}
\author{V.~V.~Nesvizhevsky}
 	\ead{nesvizhevsky@ill.fr}
 \address{Institute Laue-Langevin, 6 rue Jules Horowitz, F-38042,
Grenoble, France}
\author{E.~V.~Lychagin, A.~Yu.~Muzychka, A.~V.~Strelkov}
 \address{Joint Institute for Nuclear Research, 141980, Dubna, Russia}
\author{G.~Pignol, K.V.~Protasov}
 \address{LPSC, UJF - CNRS/IN2P3 - INPG, 53, rue des Martyrs, F-38026 Grenoble, France}

\date{\today}

\begin{abstract}
We study possibility of efficient reflection of very cold neutrons (VCN) from powders of nanoparticles. 
In particular, we measured the scattering of VCN at a powder of
diamond nanoparticles as a function of powder sample thickness, neutron
velocity and scattering angle. 
We observed extremely intense scattering of VCN even off thin powder samples. 
This agrees qualitatively with the model of independent nanoparticles at
rest. 
We show that this intense scattering would allow us to use nanoparticle
powders very efficiently as  the very first reflectors for neutrons with
energies within a complete VCN range up to $10^{-4}$ eV. 
\end{abstract}


\end{frontmatter}

%--------------------%
\section{Introduction}
%--------------------%

In the present article we report  on the feasibility of efficiently
reflecting very cold neutrons (VCN) at a powder of diamond nanoparticles,
thus bridging the energy gap between efficient reactor reflectors \cite{Fermi} for thermal and cold neutrons, and the effective Fermi
potential for ultracold neutrons (UCN) \cite{Golub,Ignatovich,Shapiro}.

The use of nanoparticles provides a sufficiently large cross-section for coherent interaction and inhomogeneity of the moderator density 
on a spatial scale of about the neutron wavelength \cite{surfacenanoparticles2}.
Nevertheless, many neutron-nanoparticle collisions are needed in order to reflect VCN 
as they scatter preferably to straightforward direction.
A large number of collisions constrain the choice of materials: only low absorbing ones with high effective Fermi potential are appropriate.
Thus, diamond nanoparticles is an evident candidate for such VCN reflector.

The formation of diamond particles by explosive shock was first observed
more than forty years ago \cite{Carli}. 
Since then very intensive studies of their production and of their various
applications have been performed worldwide. These particles measure a few
nanometers \cite{ultradiamond}. Nanoparticles consist of a diamond nucleus
within an onion-like shell with a complex chemical composition
\cite{Aleksenskii}. 
A recent review of the synthesis, structure, properties and applications of
diamond nanoparticles can be found in~\cite{Dolmatov}.

Scattering off nanoparticles could be used also to study the dynamics of
nanoparticles, at a surface, in the material bulk
\cite{surfacenanoparticles1,originOfSmallEnergyChanges,surfacenanoparticles3}, 
and even probably to cool VCN down to the UCN energy range
\cite{gel1,gel2,gel3}. 
On the other hand, other studies have shown that the quasi-elastic
scattering of UCN at nanoparticles on solid surfaces
could be responsible for false effects in storage experiments in
fundamental particle physics
\cite{losstrap2,losstrap3,losstrap4,surfacenanoparticles4,surfacenanoparticles5}. 

The first experiments on the reflection of VCN from nanostructured
materials as well as on VCN storage were carried out in the seventies in
\cite{Steyerl} and later continued in \cite{Arzumanov}. 
We have extended the energy range and the efficiency of VCN reflection
by exploiting diamond nanoparticles.
A reflector of this type would be particularly useful for 
both UCN sources using ultracold nanoparticles
\cite{surfacenanoparticles2,moderator} and for VCN sources; it would not be
efficient however for cold and thermal neutrons, as shown in~\cite{Artemiev}.
We estimated the maximum energy of the reflected neutrons. We also explored
the reliability of the theoretical predictions for the interaction of
neutrons with nanoparticle powders, and we provide an overview of the
phenomena relevant to the interaction.  

In section II we describe the experimental installation developed for these
studies, and the measuring procedure.
In section III we describe the samples of ultradiamond  powder used (so
called ultradiamond90 \cite{ultradiamond}) and the procedures for measuring
the intensity and the angular distribution of the scattered neutrons. 
In section IV we analyse the results, comparing them with Monte Carlo
simulations based on a model of independent nanoparticles at rest.

%-------------------------------------%
\section{The experimental installation}
%-------------------------------------%

\begin{figure}
\begin{center}
\includegraphics[width=.8\linewidth]{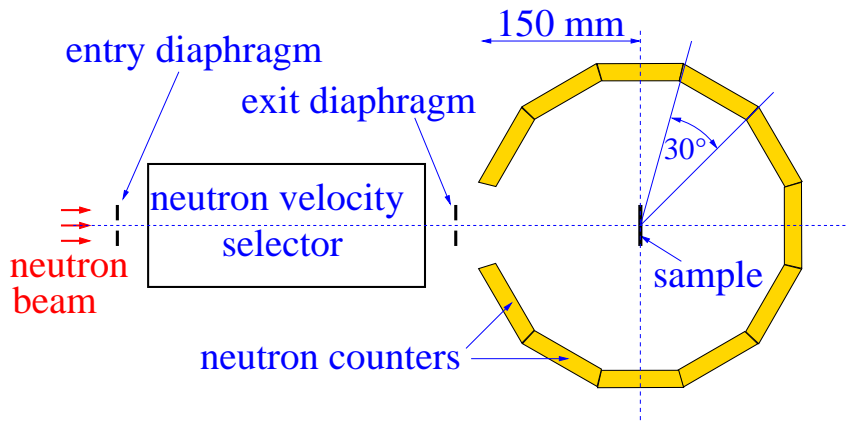}
\end{center}
\caption{The experimental setup (view from above)} \label{Setup}
\end{figure}

The setup is shown in fig. \ref{Setup} (viewed from above). 
Two boron rubber diaphragms with a thickness of 5 mm shape the neutron beam; 
the beam axis is set in a horizontal plane. 
The diameter of the entrance diaphragm is 16 mm; 
the diameter of the exit diaphragm is 10 mm; 
the distance between the diaphragms is 116 cm. 
A mechanical velocity selector is installed between them. 
A sample is placed at a distance of 23 cm downstream of the exit diaphragm
in the middle of the detector assembly. The detector assembly consists of
11 neutron counters measuring the neutrons scattered by the sample. 
The counters' flat entrance windows form 11 sides of a regular 12-sided
prism. 
The neutron beam enters through the remaining open 12$^{\rm th}$ side,
perpendicular to the assembly. 
The counter directly opposite the entrance measures the neutrons passing
through; 
its window is covered by a cadmium plate of thickness 0.5 mm and is pierced
with a 3 cm-diameter hole in the line of the beam. 
The other counters measure the neutrons scattered $30^\circ$ in a
horizontal plane and $60^\circ$ in a vertical plane 
(the solid angle of $\pi/6$). 
The detector assembly is surrounded by cadmium, borated polyethylene and
boron rubber neutron shieldings. 
In order to decrease neutron scattering in the air we placed the detector
assembly in a polyethylene bag filled with argon (Ar). The neutron entry
window consisted of a flat quartz plate with a thickness of 3.5 mm. 

Each of the 11 proportional gaseous neutron counters has a thin aluminum
window with a thickness of 200 $\mu$m. 
The counters are filled with a mixture of Ar and $^3$He gas at a pressure
of 1 bar. 
The $^3$He partial pressure is equal to 130 mbar; the thickness of the
gas-sensitive layer is 18 mm. 
The detection efficiency for neutrons with a velocity of 100 m/s is
therefore equal to $\approx 50 \%$; the efficiency for neutrons with a
velocity of 40 m/s is equal to $\approx 85 \%$. 
It should be noted that absolute efficiency values are irrelevant when
measuring neutron elastic scattering probabilities if all the counters are
equally efficient. 

 Finally, the neutron velocity selector has a resolution of 10--15 $\%$ in
the whole VCN velocity range.

%-------------------------------------------------------%
\section{The experiment with diamond powder nanoparticles}
%-------------------------------------------------------%

The measurement was carried out at the VCN PF2 beam position \cite{PF2} 
at the Institut Laue-Langevin (Grenoble, France), providing access to
neutrons with a velocity in the range of 30 - 160~m/s. 
Our sample was a powder of diamond nanoparticles
(ultradiamond90 \cite{ultradiamond}) with diameters in the range of 2--5~nm
and a known size distribution \cite{ultradiamond}.

The powder was placed between two sapphire plates with a thickness of 1 mm
each; 
the space between the plates was sealed on all sides with aluminum or
Teflon foil. 
Once the powder was in position it was subjected to vibration to increase
its density; the resulting density was equal to 0.6 g/cm$^3$. 
We prepared four samples with a thickness of 0.2, 0.4, 2 and 6 mm.
The choice of thicknesses was determined as follows.
The thinner the sample, the lower the multiple scattering effect.
Thinner samples thus allow us to study neutron-nanoparticle interactions.
The thicker the sample, the larger the reflectivity. Thicker samples thus
allow us to assess the feasability of using nanopowder as a reflector.
The minimal thickness of a layer of this type, without significant holes,
is 0.2 mm.

\begin{figure}
\begin{center}
\includegraphics[width=0.8\linewidth]{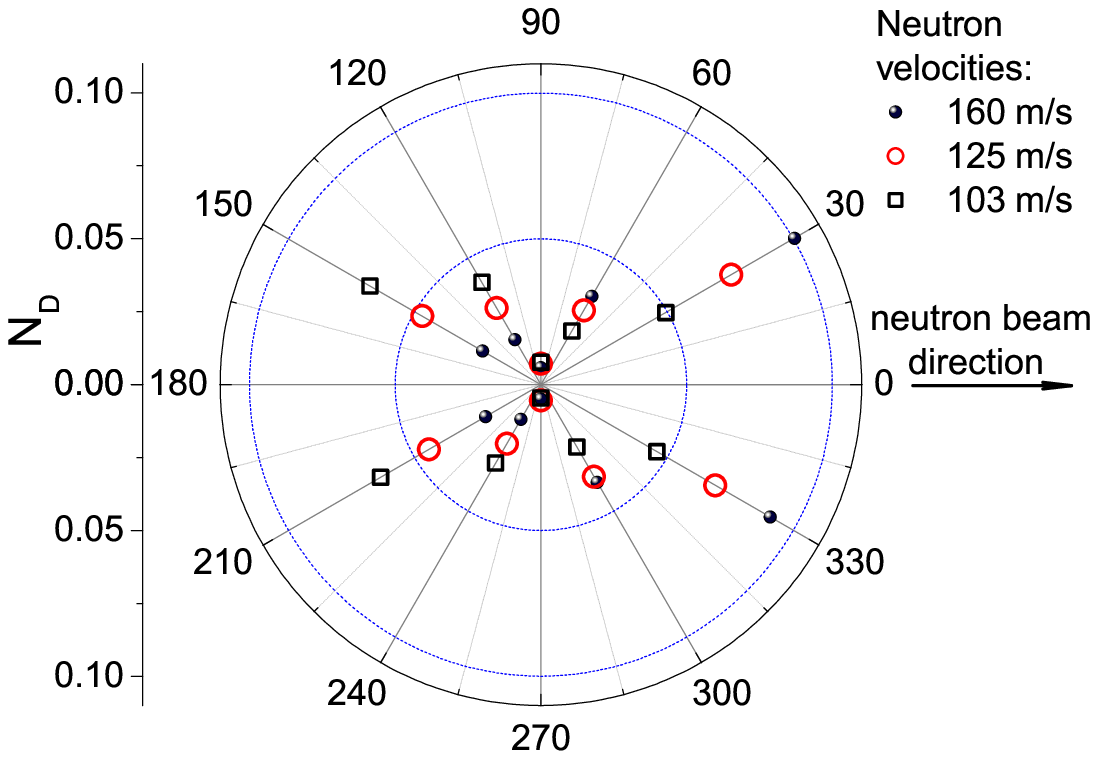}
\includegraphics[width=0.8\linewidth]{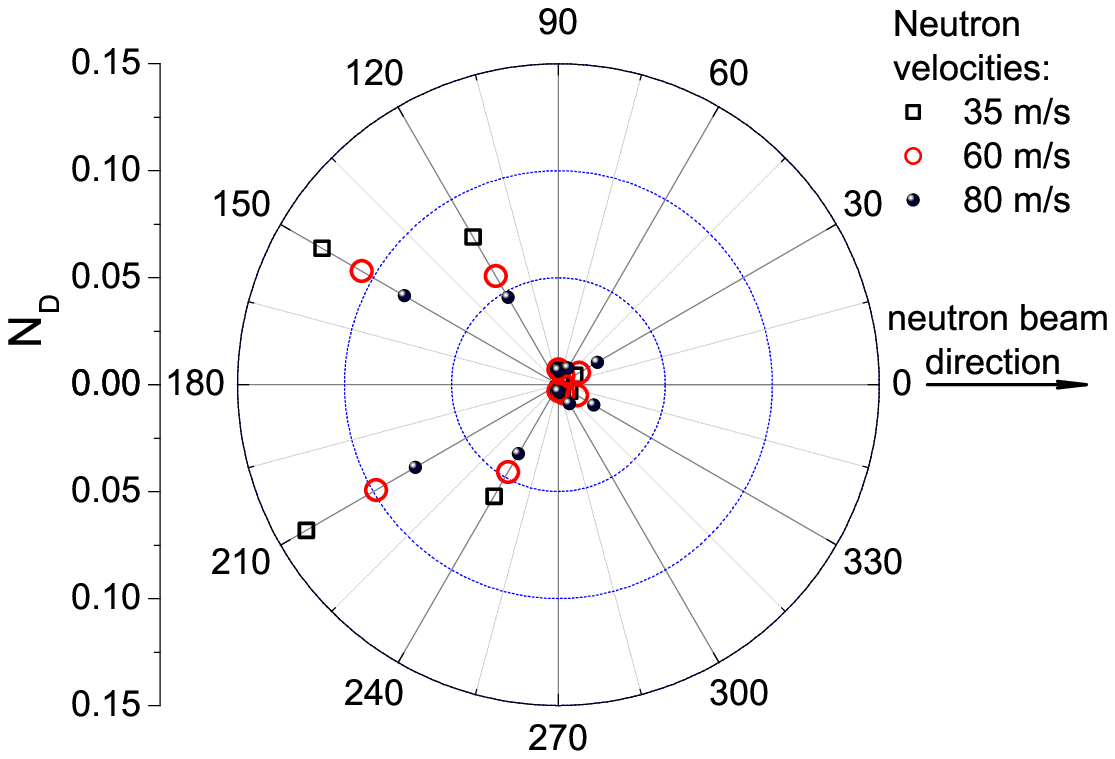}
\end{center}
\caption{
The scattered neutron flux (normalized to the initial neutron flux) 
is shown as a function of the scattering angle for an ultradiamond90 powder
sample with a thickness of 2 mm. 
 The angle measured is that between the incoming beam of neutrons and its
line of travel from the sample to the center of a corresponding counter.
The results for all measured velocities from 35~m/s to 160~m/s are
indicated by the different types of point.
}
\label{Result}
\end{figure}

Fig. \ref{Result} presents examples of the angular dependencies measured
for neutrons scattered at a 2mm-thick sample of ultradiamond90.

% ---------------- check elasticity ----------------

\begin{figure}
\begin{center}
\includegraphics[width=.8\linewidth]{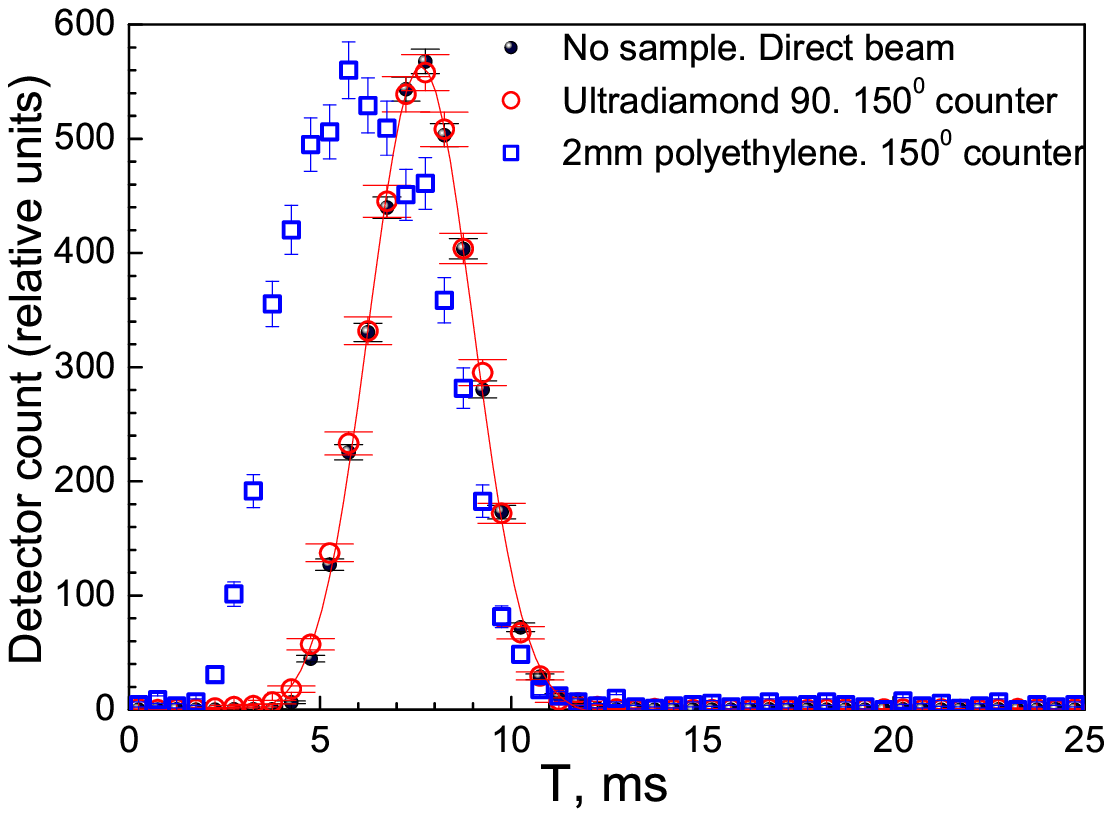}
\end{center}
\caption{
The neutron count rate is presented as a function of the time of flight of
the neutrons with an average initial velocity of 60~m/s. 
The black circles correspond to the initial neutron spectrum. 
The open circles indicate the data for the spectrum of neutrons scattered
to an angle of $150^\circ$. 
The thickness of the ultradiamond90 powder sample is equal to 2 mm. 
The squares show results for the scattering of neutrons at a polyethylene
sample with a thickness of 2 mm, measured at the same counter.
} 
\label{tof}
\end{figure}

In order to check the elasticity of the neutron scattering at the
ultradiamond90 powder samples, 
we measured the time-of-flight spectrum of the scattered neutrons using a
counter placed at an angle $150^\circ$. 
The average initial neutron velocity was equal to 60 m/s. 
The results of a comparison of the spectrum of scattered neutrons with the
initial spectrum measured in analogous fashion are presented in fig.
\ref{tof}. 
To demonstrate the sensitivity of this method, we also measured the
scattering of neutrons at a polyethylene sample with a thickness of 2 mm.
We were thus able to show that VCN scattering at the nanoparticles is
mainly elastic.

%----------------%
\section{Analysis}
%----------------%

In order to analyse the data measured we developed a Monte Carlo model of
the experiment.
The interaction of the neutrons with a nanoparticle powder can be described
using the simple approach proposed in \cite{moderator}.
We neglected the relatively complex internal structure of the nanoparticle
and modelled it as a uniform sphere. 
The neutron-nanoparticle elementary interaction was calculated using
the first Born approximation.
The absorption and elastic cross-sections, as well as the angular
distribution of the scattered neutrons, were calculated analytically. 

The chemical composition of the nanoparticle is complex and includes carbon
(up to 88\%), hydrogen (1.0\%), nitrogen (2.5\%), oxygen (up to 10\%)
\cite{Vereschagin}. Moreover, a certain amount of water covers a
significant surface area of the nanoparticles. 
 In general, the hydrogen in the water and on the surface of the
nanoparticles scatters the neutrons up to the thermal energy range
("up-scattering''); 
thermal neutrons do not interact as efficiently with nanoparticles and
therefore traverse the powder. 
As the amount of hydrogen was not precisely known  we considered it as a
free parameter in the model. 

We built a Monte Carlo model of the experiment on this basis, taking into
account the sample and installation geometry described above. 
The size distribution of the nanoparticles in the powder is assumed to be
equivalent to that presented in \cite{ultradiamond}.

\begin{figure}
\begin{center}
\includegraphics[width=.65\linewidth]{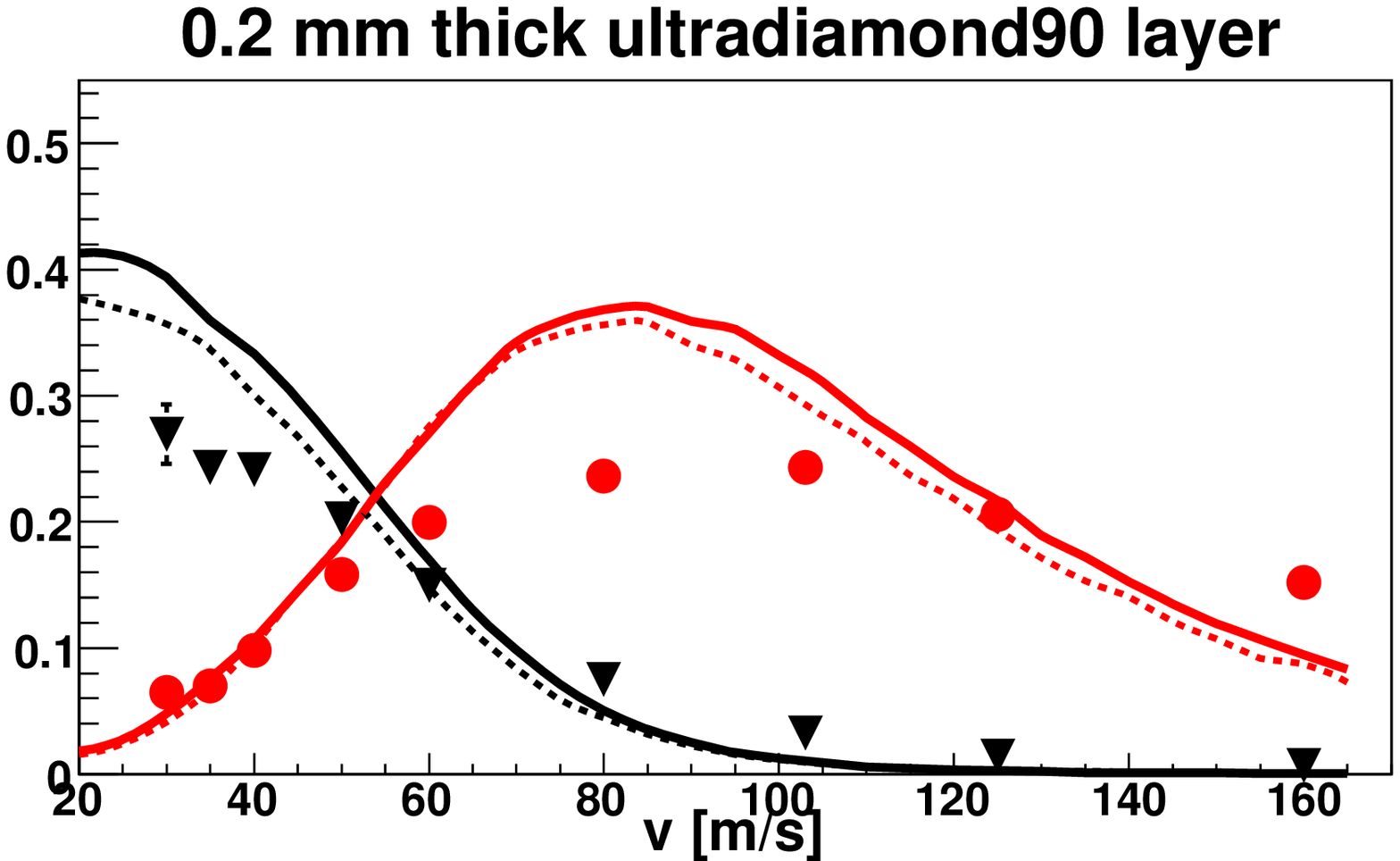}
\includegraphics[width=.65\linewidth]{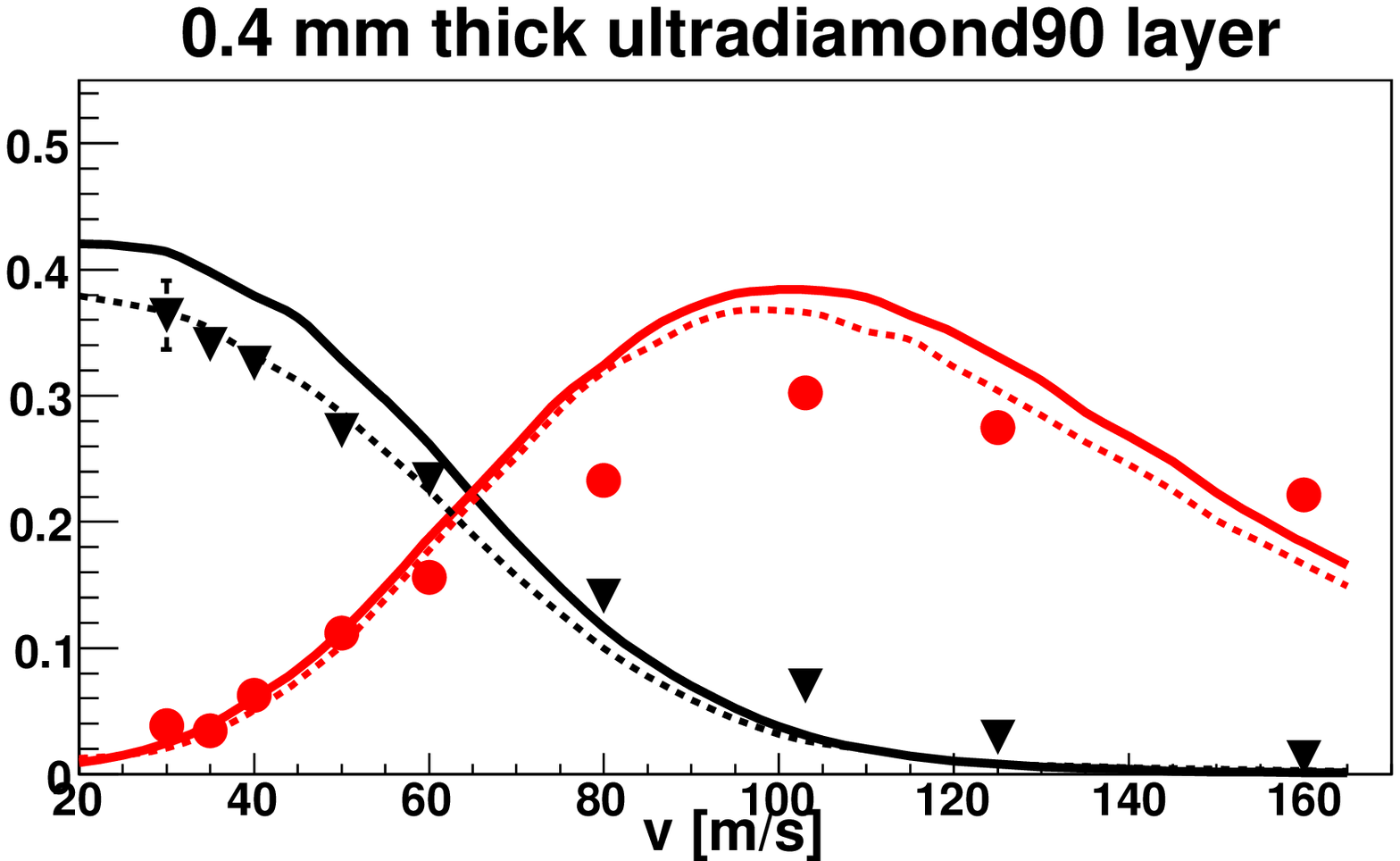}
\includegraphics[width=.65\linewidth]{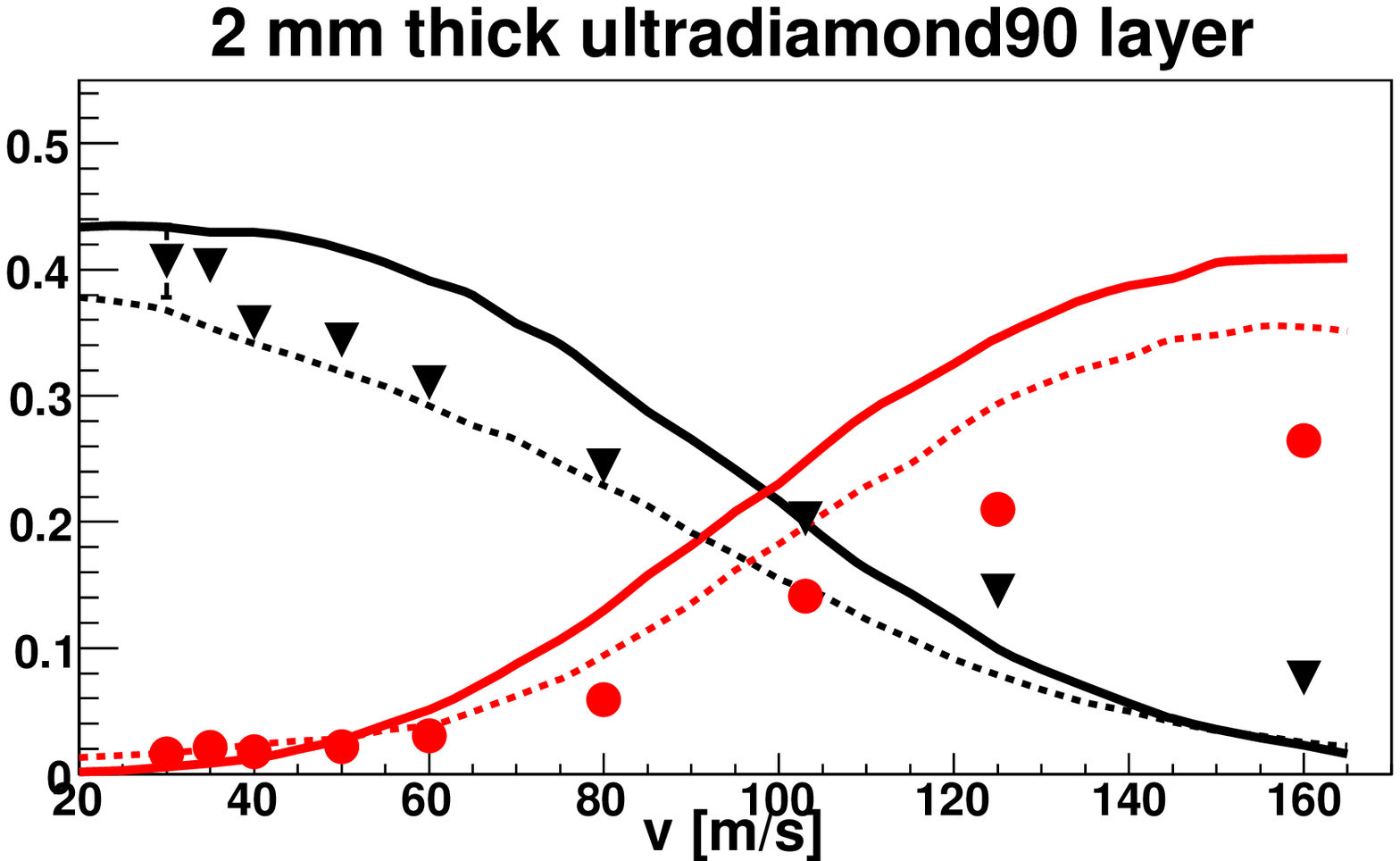}
\includegraphics[width=.65\linewidth]{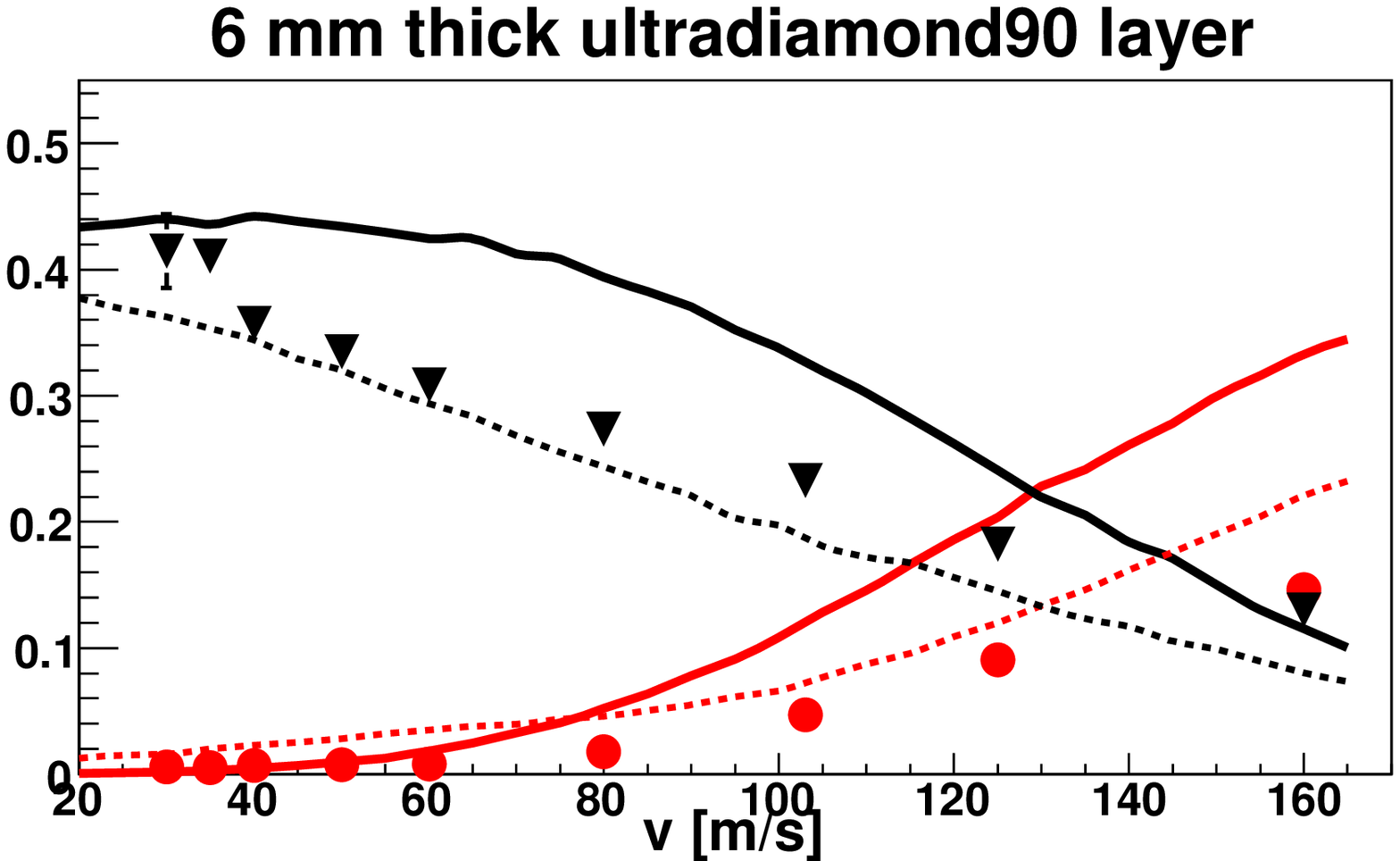}
\end{center}
\caption{
The detection probability dependencies  for backward (counters at
120$^\circ$ + 150$^\circ$ + 210$^\circ$ + 240$^\circ$) and forward
(counters at 30$^\circ$ + 60$^\circ$ + 300$^\circ$ + 330$^\circ$)
scattering of VCN at the ultradiamond90 powder are shown as a function of
VCN velocity, for different powder thicknesses.
The triangles correspond to the measured probability of VCN backward
scattering. The circles indicate the measured probability of VCN forward
scattering.
The solid lines correspond to the model of independent nanoparticles at rest.
The dotted lines indicate the model that takes into account the
up-scattering of VCN at water in the powder.
} 
\label{DataMonteCarlo}
\end{figure}

The simulation results are compared to the probabilities measured of
forward and backward scattering of VCN at samples of all measured
thicknesses as shown in fig. \ref{DataMonteCarlo}. 
The two sets of curves correspond to the calculation with pure carbon
nanoparticles (solid lines) and to that with 1\% of hydrogen added (dotted
lines).

In the broad range of VCN velocities and sample thicknesses, good
qualitative agreement was obtained between the probability and the angular
distribution of scattered neutrons, 
and the predictions of the model of independent nanoparticles at rest.
The hydrogen in the powder suppresses the penetration of neutrons across
the samples more strongly than it decreases their reflection; this effect
is more pronounced for small neutron velocities and for thick samples. 
In future studies we will remove the water and other hydrogen-containing
impurities from the samples and cool the powder down to liquid nitrogen
temperature (at which the upscattering of neutrons at hydrogen is largely
suppressed).
The residual discrepancy is probably due to poor knowledge of the spatial
density distribution in the powder. This variation and even holes in the
samples could affect the measurements, for the thinnest samples in particular.

\begin{figure}
\begin{center}
\includegraphics[width=.95\linewidth]{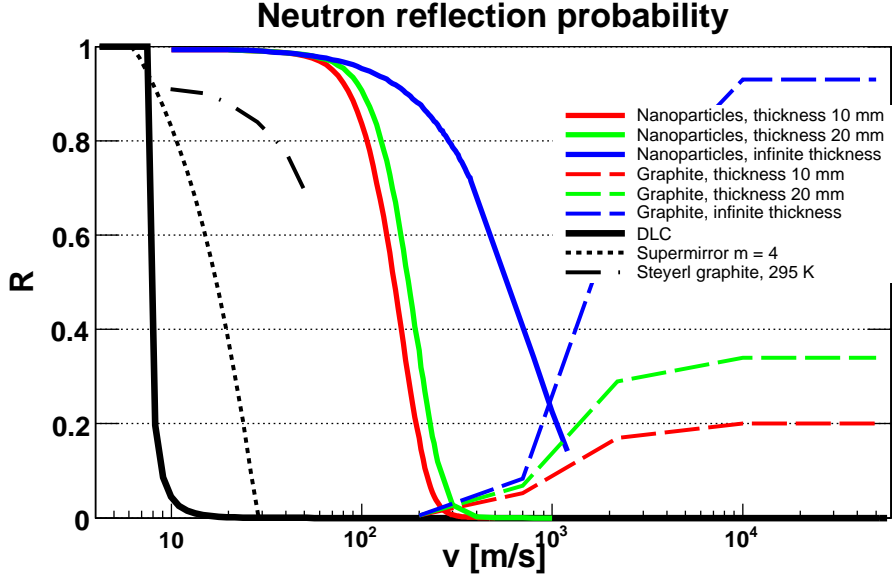}
\end{center}
\caption{
The neutron reflection probability is shown as a function of neutron velocity for various carbon-based reflectors:
1) Diamond-like coating (DLC), with the highest known value of Fermi-potential (thick solid line) 
2) The best present supermirror \cite{supermirrors} (dotted line) 
3) Steyerl's measurement of VCN reflection \cite{Steyerl} (dased-dotted line) 
4) Hydrogen-free ultradiamond90 powder for thicknesses of 10 mm, 20 mm, and infinite (solid lines)
5) Reactor graphite reflectors at room temperature for thicknesses of 10 mm, 20 mm, and infinite (dashed lines).
} \label{PredictionMonteCarlo}
\end{figure}

As can be  seen from fig.~4, the maximum energy of the reflected VCN and
the reflection probability far exceed the corresponding values for the best
supermirrors available \cite{supermirrors}, although the reflection is not
specular at the nanopowder reflector of course.

If the reflection of the neutrons from the ultradiamond90 powder is
determined uniquely by their scattering at the nanoparticles, and if we
eliminate any additional losses (due in particular to hydrogen-containing
impurities), the reflection probability would be as high as is shown in
fig. \ref{PredictionMonteCarlo}.
Efficient reflection would enable VCN to be stored in closed traps.
We intend to explore this exciting hypothesis experimentally.

%------------------%
\section{Conclusion}
%------------------%

We have observed extremely intense scattering of VCN at diamond powder
nanoparticles for the first time.
We have demonstrated that nanoparticle powders can be efficient reflectors
of VCN with an energy value as high as $10^{-4}$ eV.
This energy value and the reflection probability far exceed values for the
best available supermirrors, although the reflection at nanoparticles is
not specular.
This phenomenon has a number of applications, including the storage of VCN
in closed traps, reflectors for VCN and UCN sources, the more efficient
guiding of VCN and, probably, of even faster neutrons.

%-----------------------%
\section{Aknowledgements}
%-----------------------%

We are grateful to Y. Calzavara, P.~Geltenbort and C.~Plonka for their assistance.
This work is supported by RFBR grant number 03-02-16784-a.

%-------------------------%

\end{document}